\documentclass[fleqn,10pt]{wlscirep}
\usepackage{amsmath}
\usepackage{graphicx}
\usepackage{multirow}
\title{Unraveling the hardness of a borophene-based compound}

\author{N. Gonzalez Szwacki}
\affil{Institute of Theoretical Physics, Faculty of Physics, University of Warsaw, ul. Pasteura 5, PL-02-093 Warsaw, Poland} 
\affil{gonz@fuw.edu.pl}


\begin{abstract}
Two-dimensional systems have strengthened their position as one of the key materials for novel applications. Very recently, boron joined the distinguished group of elements that are confirmed to possess 2D allotropes, named borophenes. In this work, we explore the stability and hardness of the highest boride of tungsten, which is regarded as built of borophenes separated by metal atoms. We show that WB$_{3+x}$ has Vickers hardnesses approaching 40~GPa only for small values of $x$. The insertion of extra boron atoms is, in general, detrimental for WB$_{3}$ in terms of hardness since leads to the formation of quasi-planar boron sheets that are less tightly connected with the adjacent W layers. Very high concentrations of boron ($x \approx 1$), give rise to a soft (Vickers hardness of about 8~GPa) and unstable $hP$20-WB$_{4}$ structure that can be considered as built of quasi-planar boron $\alpha$-sheets separated by graphitic W layers. On the other hand, we show that the formation of tungsten vacancies brings on structures, e.g. W$_{0.75}$B$_{3+x}$, with Vickers hardnesses that are less sensitive to variations in the boron content and are close in value to the experimentally reported load-independent values above 20~GPa.
\end{abstract}
\begin{document}

\flushbottom
\maketitle
\thispagestyle{empty}

\section*{Introduction}
The highest boride of tungsten -- often referred to as tungsten tetraboride -- is recently best explored for its potential applications as superhard material, however made its first appearance in the literature in 1961, when Chretien and Helgorsky\cite{chretien1961} did the first attempt to find its structure. Years later Romans and Krug \cite{romans1966} reported that WB$_{4}$ has a hexagonal structure of 20 atoms per unit cell and lattice constants of 5.2 and 6.34 {\AA} for $a$ and $c$, respectively. The space group of this structure was determined to be $P6_{3}/mmc$. The $hP$20-WB$_{4}$ structure serves now as a reference structure for almost all subsequent experimental studies related to boron-rich materials with a WB$_{4}$-like structure.\cite{gu2008,liu2011,mohammadi2011,xie2012,xiong2013} The mechanical properties of WB$_{4}$ were first determined by Gu \textit{et al.}\cite{gu2008} who reported Vickers hardness ($H_{\text{V}}$) values of 46.2 and 31.8 GPa under applied loads of 0.49 and 4.90 N, respectively, measured by the microindentation technic. Subsequently, Mohammadi \textit{et al.} \cite{mohammadi2011} also measured the hardness by microindentation method and reported $H_{\text{V}}$ values of 43.3 and 28.1 GPa at low (0.49~N) and high (4.90~N) loads, respectively. More recently, the Vickers hardness for W$_{0.85}$B$_{3}$ was reported by Tao \textit{et al.} \cite{tao2014} to be 42.0 and 25.5 GPa under applied loads of 0.098 and 4.90 N, respectively. Finally, Lech \textit{et al.} \cite{lech2015} determined the maximum nanoindentation hardness of W$_{0.82}$B$_{3.54}$ (at a penetration depth of 95.25~nm) to be 41.7~GPa. It is generally accepted that a reliable hardness of a material can be determined from the asymptotic hardness region achieved at high loads.\cite{tian2012} The quite large differences between the $H_{\text{V}}$ values reported for WB$_{4}$, especially for high loads, can be attributed to differences in the amount of boron contamination and/or presence of tungsten vacancies, which were experimentally seen in the studied samples.\cite{lech2015} By exploring structures with different compositions, we are able to explain on the theoretical ground the apparent differences between the reported experimental results.

\begin{figure}[h]
\centering
  \includegraphics[width=.5\linewidth]{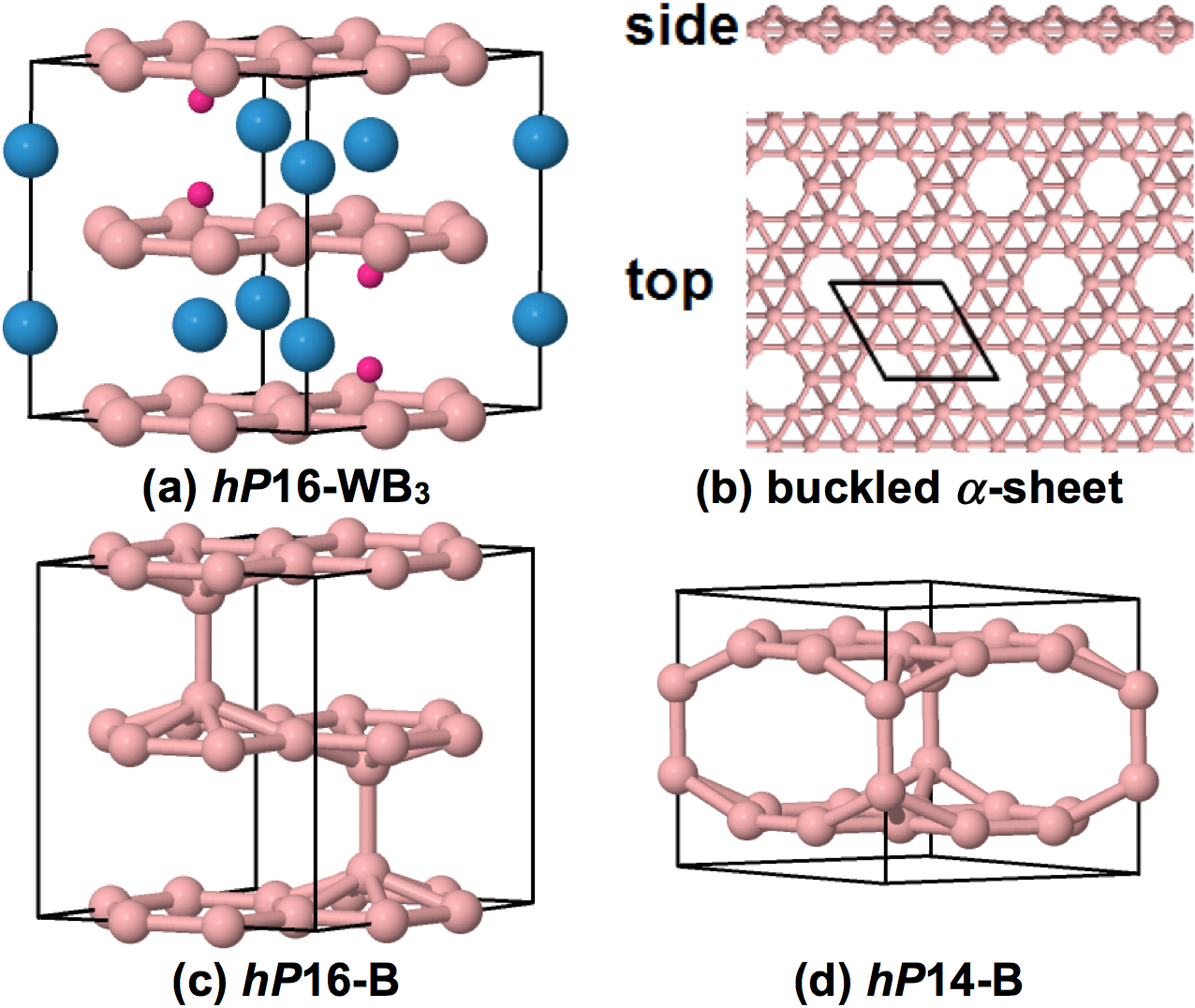}
  \caption{(Color online). (a) The $hP$16-WB$_{3}$ structure. The large and small spheres represent W and B atoms, respectively. The small red dots indicate the position of the 4 extra boron atoms that are present in the $hP$20-WB$_{4}$ structure. (b) Buckled boron $\alpha$-sheet present in the $hP$20-WB$_{4}$ structure. (c) The $hP$16-B structure that is obtained by removing all the tungsten atoms from $hP$20-WB$_{3}$. (d) The $hP$14-B structure that derives from $hP$16-B by removing one of the two dimers.}
\label{fig1}
\end{figure}

The common description of $hP$20-WB$_{4}$ that can be found in the literature is that this structure consists of graphitic boron layers separated by graphitic layers of W atoms like in the $hP$16-WB$_{3}$ structure but with additional B$_2$ dimers located between boron sheets and aligned along the $c$-axis (see Fig.~\ref{fig1}a). This description, although very elegant, is completely decoupled from more recent investigations related to 2D boron crystals.\cite{tang2007,feng2016} An `updated' view to $hP$20-WB$_{4}$ would be that it is a structure consisting of a sequence of quasi-planar boron $\alpha$-sheets separated by graphitic W layers. Extensive theoretical investigations have proved, however, that the stoichiometric $hP$20-WB$_{4}$ structure is thermodynamically and dynamically unstable. Its calculated enthalpy of formation is positive, with a value of 0.4~eV/atom,\cite{cheng2014} and from its phonon dispersion the structure was shown to be highly unstable.\cite{zhang2012} In fact, we argue that the formation of stable quasi-planar boron layers within the $hP$20-WB$_{4}$ structure is the main reason for the instability of this and related boron reach structures. Therefore $hP$20-WB$_{4}$ is in that view a nonexistent structure.

In more recent reports,\cite{cheng2013, cheng2014} the highest borides of tungsten are described as $hP$16-WB$_{3}$ structures contaminated with additional boron atoms. The exact position of the boron atoms in the crystal lattice are difficult to be determined experimentally because of the large mass difference between W and B atoms.\cite{cheng2013} Therefore, the combination of theory and experiment is essential for the understanding of the observed findings. Since in the experiment the $hP$16-WB$_{3}$ structure is not only contaminated with boron atoms but also, to some extent, possesses tungsten vacancies,\cite{zeiringer2014, tao2014, lech2015} in this work a more precise notation is used when referring to the highest boride of tungsten, namely W$_{1-y}$B$_{3+x}$, to underline the existence of W vacancies and explore their influence on the stability and properties of WB$_{3+x}$.

\begin{figure}[h]
\centering
  \includegraphics[width=.48\linewidth]{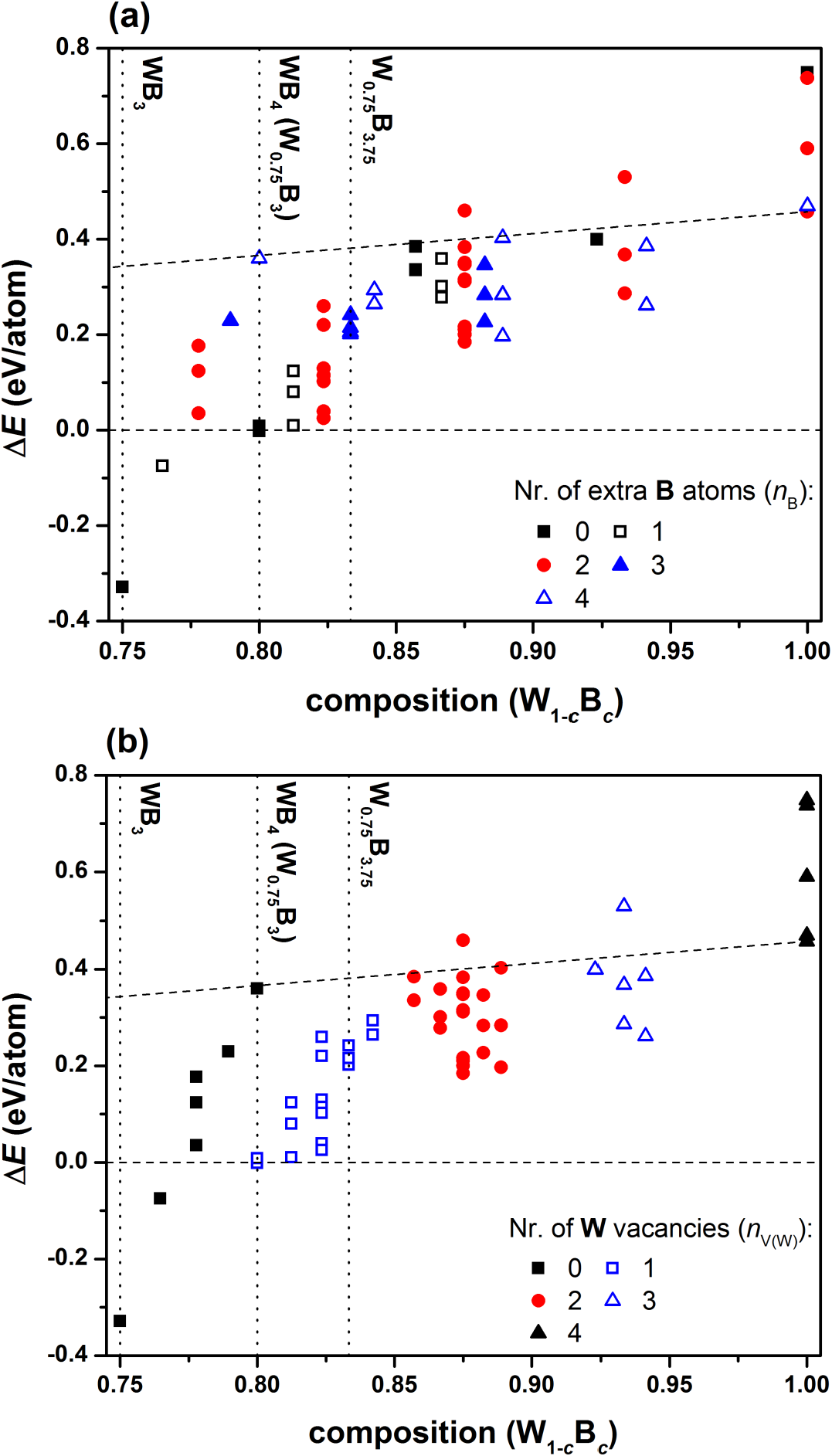}
  \caption{(Color online). Enthalpies of formation for structures derived from $hP$16-WB$_{3}$ by selective removal of W atoms and/or contamination by additional B atoms. The vertical dotted lines correspond to compositions that were reported in the literature.}
\label{fig2}
\end{figure}
\begin{figure}[h]
\centering
  \includegraphics[width=.48\linewidth]{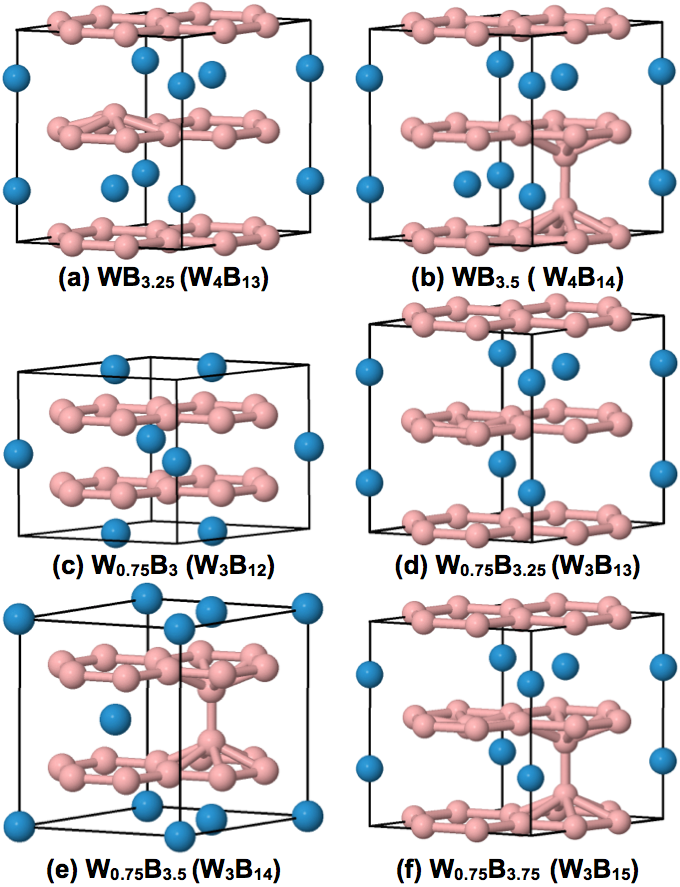}
  \caption{(Color online). The crystal structure of several boron reach phases of the W-B system. The large and small spheres represent W and B atoms, respectively. The W$_{m}$B$_{n}$ notation in parenthesis shows the number of W and B atoms in the unit cell.}
\label{fig3}
\end{figure}
\begin{figure}[h]
\centering
  \includegraphics[width=.48\linewidth]{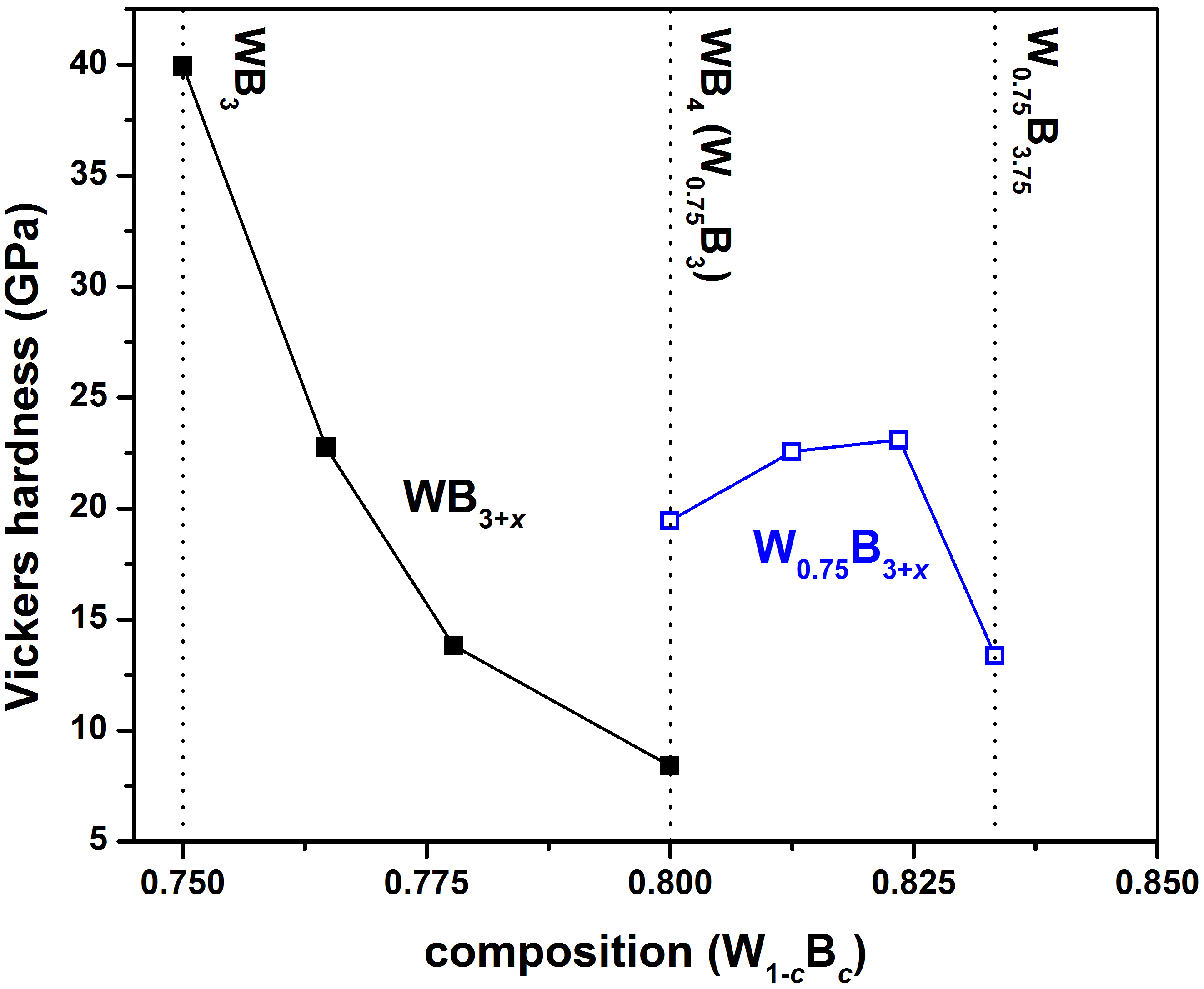}
  \caption{(Color online). Vickers hardnesses calculated for the most stable structures at a given composition. The filled and open squares correspond to structures of the WB$_{3+x}$ and W$_{0.75}$B$_{3+x}$ type, respectively.}
  \label{fig4}
\end{figure}
\begin{table}[h]
\centering
\caption{Phases, occupations relative to the $P6_{3}/mmc$ space group, lattice constants, and resulting space groups of the most stable structures at a given composition.}
\label{my-label}
\begin{tabular}{lccccccc}
\hline
\multirow{4}{*}{Phase}                                            

& \multicolumn{4}{c}{\begin{tabular}[c]{@{}c@{}}Occ. relative to\\ $P6_{3}/mmc$ \end{tabular}}                                                                                                                             
& \multirow{4}{*}{$a$ ({\AA})} & \multirow{4}{*}{$c$ (\AA)} & \multirow{4}{*}{\begin{tabular}[c]{@{}c@{}}Resulting \\ space group\end{tabular}} \\ \cline{2-5}
                                                             & \begin{tabular}[c]{@{}c@{}}W1 \\ ($2c$)\end{tabular} & \begin{tabular}[c]{@{}c@{}}W2 \\ ($2b$)\end{tabular} & \begin{tabular}[c]{@{}c@{}}B1 \\ ($12i$)\end{tabular} & \begin{tabular}[c]{@{}c@{}}B2 \\ ($4f$)\end{tabular} &                        &                        &                                                                                \\ \hline
\begin{tabular}[c]{@{}l@{}}WB$_{3}$\end{tabular}        & 1                                                  & 1                                                  & 1                                                   & 0                                                  & 5.171                 & 6.228                 & \begin{tabular}[c]{@{}c@{}}$P6_{3}/mmc$\end{tabular}                        \\
\begin{tabular}[c]{@{}l@{}}WB$_{3.25}$\end{tabular}     & 1                                                  & 1                                                  & 1                                                   & 0.25                                               & 5.173                 & 6.405                 & \begin{tabular}[c]{@{}c@{}}$P3m1$\end{tabular}                           \\
\begin{tabular}[c]{@{}l@{}}WB$_{3.5}$\end{tabular}      & 1                                                  & 1                                                  & 1                                                   & 0.5                                               & 5.195                 & 6.398                 & \begin{tabular}[c]{@{}c@{}}$P3m1$\end{tabular}                           \\
\begin{tabular}[c]{@{}l@{}}WB$_{4}$\end{tabular}        & 1                                                  & 1                                                  & 1                                                   & 1                                                  & 5.327                 & 6.334                 & \begin{tabular}[c]{@{}c@{}}$P6_{3}/mmc$\end{tabular}                        \\
\begin{tabular}[c]{@{}l@{}}W$_{0.75}$B$_{3}$\end{tabular}    & 1                                                  & 0.5                                                & 1                                                   & 0                                                  & 5.416                 & 5.271                 & \begin{tabular}[c]{@{}c@{}}$P6/mmm$\end{tabular}                         \\
\begin{tabular}[c]{@{}l@{}}W$_{0.75}$B$_{3.25}$\end{tabular} & 0.5                                                & 1                                                  & 1                                                   & 0.25                                               & 5.144                 & 6.305                 & \begin{tabular}[c]{@{}c@{}}$P3m1$\end{tabular}                           \\
\begin{tabular}[c]{@{}l@{}}W$_{0.75}$B$_{3.5}$\end{tabular}  & 1                                                  & 0.5                                                & 1                                                   & 0.5                                               & 5.139                 & 6.398                 & \begin{tabular}[c]{@{}c@{}}$P\bar{6}m2$\end{tabular}                          \\
\begin{tabular}[c]{@{}l@{}}W$_{0.75}$B$_{3.75}$\end{tabular} & 1                                                  & 0.5                                                & 1                                                   & 0.75                                               & 5.211                 & 6.291                 & \begin{tabular}[c]{@{}c@{}}$P3m1$\end{tabular} \\ \hline                       
\end{tabular}
\end{table}
\begin{table}[h]
\centering
\caption{Number of W vacancies $n_{\text{V(W)}}$, number of extra B atoms $n_{\text{B}}$, bulk modulus $B$ (GPa), shear modulus $G$ (GPa), Young's modulus $E$ (GPa), Poisson's ratio $\nu$, and Vickers hardness $H_{\text{V}}$ (GPa) for several W$_{1-y}$B$_{3+x}$ structures.}
\label{my-label}
\begin{tabular}{lccccccc}
\hline
\multicolumn{1}{l}{Phase} & $n_{\text{V(W)}}$ & $n_{\text{B}}$ & $B$ & $G$ & $E$ & $\nu$ & $H_{\text{V}}$ \\ \hline
WB$_{3}$ & 0 & 0 & 315 & 266 & 622 & 0.17 & 39.9 \\
WB$_{3.25}$ & 0 & 1 & 283 & 185 & 456 & 0.23 & 22.8 \\
WB$_{3.5}$ & 0 & 2 & 317 & 156 & 403 & 0.29 & 13.8 \\
WB$_{4}$ & 0 & 4 & 321 & 126 & 335 & 0.33 & 8.4 \\
W$_{0.75}$B$_{3}$ & 1 & 0 & 289 & 173 & 433 & 0.25 & 19.5 \\
W$_{0.75}$B$_{3.25}$ & 1 & 1 & 254 & 171 & 419 & 0.22 & 22.6 \\
W$_{0.75}$B$_{3.5}$ & 1 & 2 & 279 & 184 & 453 & 0.23 & 23.1 \\
W$_{0.75}$B$_{3.75}$ & 1 & 3 & 249 & 131 & 334 & 0.28 & 13.4 \\ \hline
\end{tabular}
\end{table}

\section*{Results}
\textbf{The structure of W$_{1-y}$B$_{3+x}$.} The highest boride of tungsten are obtained starting from $hP$16-WB$_{3}$ by adding additional boron atoms at the positions shown in red in Fig.~1a and/or by selective removal of W atoms. The fully `packed' structure is the $hP$20-WB$_{4}$ structure that has buckled boron $\alpha$-sheets separated by W layers. The buckling height is 1.49~{\AA} and is larger than that of the freestanding triangular boron sheet (0.82~{\AA} \cite{kunstmann2006}).  The complete removal of all the W atoms from $hP$20-WB$_{4}$ leads to the all-boron structure that consists of 16 atoms per unit cell and is shown in Fig.~1c. This structure is nothing more than a sequence of quasi-planar boron $\alpha$-sheets arranged in such a way that the boron atoms that stick out of the graphitic frame face each other forming dimers. In the same way, we can describe the $hP$20-WB$_{4}$ structure but this time the boron $\alpha$-sheets are separated by W graphitic layers. The $hP$16-B structure is less stable than the $\alpha$-rhombohedral boron ($hR12$-B) by 0.47~eV/atom. Interestingly enough, by removing one of the boron dimers in $hP$16-B a slightly more stable (by 12~meV/atom) structure shown in Fig.~1c is obtained. The $hP$16-B and $hP$14-B structures have $P6_{3}/mmc$ and $P6/mmm$ space groups, respectively, and $a=5.034$~{\AA}, $c=6.166$~{\AA} and $a=5.081$~{\AA}, $c=5.195$~{\AA} lattice constants, respectively.

\textbf{Stability of the compounds.} To explore the relative stability of the generated structures, we calculate for each structure its enthalpy of formation per atom, $\Delta$$E$. The $\Delta$$E$ values are calculated relative to the chemical potentials of W and B atoms obtained based on the body-centered cubic tungsten ($cI2$-W) and $\alpha$-rhombohedral boron ($hR12$-B), respectively. The results of $\Delta$$E$ for W$_{1-c}$B$_{c}$ versus composition $c$ are summarized in Figs. 2a and 2b. All the structures that have enthalpies of formation above the horizontal dashed lines in Figs. 2a and 2b are, in principle, thermodynamically unstable. It is instructive, however, to draw also a line that connects $cI2$-W with $hP$14-B, what is shown in Figs. 2a and 2b by a dashed line that is above the horizontal dashed line. The enthalpies of formation of almost all the considered structures W$_{1-y}$B$_{3+x}$ are located bellow the $cI2$-W~$\leftrightarrow$~$hP$14-B line. This means that the incorporation of W atoms in between boron sheets is energetically favourable. The enthalpies of formation versus composition are presented in two ways. In Fig.~2a, we organized the results according to the number of extra boron atoms, $n_{\text{B}}$, in W$_{1-y}$B$_{3+x}$ relative to $hP$16-WB$_{3}$. In Fig.~2b, the same results have been organized emphasizing the number of W vacancies in W$_{1-y}$B$_{3+x}$ also relative to $hP$16-WB$_{3}$. It is clear from Fig.~2a that negative enthalpies of formation or positive $\Delta$$E$ but close to 0, have structures with none or no more than 2 extra B atoms. The cases with 3 and 4 extra B atoms have $\Delta$$E\simeq0.2$~eV/atom or larger. This also includes the highly debated $hP$20-WB$_{4}$ structure ($\Delta$$E=0.36$~eV/atom). From Fig.~2b, we can learn that the only relevant cases are those for which the number of W vacancies, $n_{\text{V(W)}}$, is 0 or 1, since among those cases we can find structures with negative or close to 0 enthalpies of formation. Combining all the information coming from Fig.~2, we choose 8 structures that in principle can be important to understand experimental results. Six of those structures are shown in Fig.~3. For each relevant composition $c$, we choose the structure with the lowest enthalpy of formation. The relevant structures (the highest boride of tungsten) are those with $c$ ranging from 0.75 (WB$_{3}$) to 0.83 (W$_{0.75}$B$_{3.75}$). The highest $c$ is chosen following Ref.~\cite{lech2015}.

The lattice constants and symmetry of the structures shown in Fig.~3 are summarized in Tab.~I. In this table, we also include, for each structure, the occupations of the B and W atoms relative to the $P6_{3}/mmc$ space group. It is important to notice that if we exclude the case of W$_{0.75}$B$_{3}$ (that appears not to match the others) and take the average of all the rest lattice constants listed in Tab.~I, we get 5.194 and 6.337 {\AA} for $a$ and $c$, respectively, that is, values that match very well those reported in the experiment (5.2 and 6.34 {\AA} for $a$ and $c$, respectively \cite{lech2015}). This may suggest that in the experiment is observed a non-stoichiometric structure with a random distribution of both the extra boron atoms and W vacancies and may farther explain the difficulties in the interpretations of X-ray and neutron diffraction data.\cite{lech2015}

\textbf{Mechanical properties of the compounds.} The elastic properties of the studied structures are summarized in Tab.~II, whereas the plot of the Vickers hardness versus composition is shown in Fig.~4. From this figure, we see that only stoichiometric WB$_{3}$ ($hP$16-WB$_{3}$) can be considered as superhard material. The hardness is however affected by contamination by extra B atoms. This is clearly seen in Fig.~4 for WB$_{3+x}$, for which the Vickers hardness changes from $\sim$40 to $\sim$8~GPa for an increase of the B content by 5\%. For the tungsten-deficient structures W$_{0.75}$B$_{3+x}$ the picture is different, namely, we obtain smaller variations of the Vickers hardness with the increase of B content (see Fig.~4). Most of the considered structures have Vickers hardnesses larger or equal to 20~GPa, what means that the highest boride of tungsten is a hard material but not superhard (at least in the range of considered compositions). A particularly soft structure is $hP$20-WB$_{4}$, which has a comparable bulk modulus to that of $hP$16-WB$_{3}$ but much smaller shear modulus (see Tab.~II). The softening of WB$_{4}$ may be attributed to the formation of stable 2D boron layers ($\alpha$-sheets) which are less tightly bound to the tungsten layers. The average nearest neighbour W--B distance is 2.324 and 2.383 {\AA} in WB$_{3}$ and WB$_{4}$, respectively, what reflects the weakening of the W--B bond in WB$_{4}$ with respect to WB$_{3}$.

In summery, we show that the insertion of extra boron atoms into the WB$_{3}$ structure is, in general, energetically unfavourable and lowers its shear modulus while keeping a high value for the bulk modulus, what effectively brings to a softer material. A high degree of boron contamination leads to the formation of quasi-planar boron $\alpha$-sheets separated by graphitic W layers in WB$_{4}$. More stable and harder than WB$_{4}$ are structures of the W$_{1-y}$B$_{3+x}$ type, in which boron contamination is accompanied by a presence of tungsten vacancies. Finally, the formation of tungsten vacancies gives rise to structures (e.g. W$_{0.75}$B$_{3+x}$) with Vickers hardnesses that are less sensitive to variations in the boron content and are close in value to the experimentally reported load-independent values above 20~GPa. Our results should provide guidance for the experiment in the design of new ways of WB$_{4}$ synthesis.

\section*{Methods}
Our first principles calculations are based on density functional theory (DFT) and the projector augmented wave (PAW) method as implemented in the Quantum ESPRESSO simulation package.\cite{qe2009} For the exchange and correlation functional, we use a revised Perdew-Burke-Ernzerhof spin-polarized generalized gradient approximation (PBEsol-GGA) functional. The plane-wave basis set is converged using a 40~Ry energy cutoff. A $8\times 8\times 8$ \textbf{\textit{k}}-point mesh and a Gaussian smearing of 0.005~Ry is used in the Brillouin Zone integration. The calculations are done using supercells containing up to 20 atoms. For each considered structure, we do a full atomic position and lattice parameter relaxation. 

A total of 60 low-energy W$_{1-c}$B$_{c}$ structures with high boron content $c$ are selected by using the cluster-expansion method implemented in the Alloy-Theoretic Automated Toolkit (ATAT).\cite{walle2002} The elastic properties of the most stable structures are calculated using the ElaStic code.\cite{golesorkhtabar2013} To compute the Vickers hardness, we employ the semi-empirical hardness model proposed by Chen \textit{et al.} \cite{chen2011} that correlates hardness with the elastic properties of the material. According to this model the expression for hardness is $H_{\text{V}}=2(k^{2}G)^{0.585}-3$, where $G$ and $k$ are the shear modulus and Pugh modulus ratio ($k=G/B$, $B=\text{bulk modulus}$), respectively.

\bibliography{v1.bib}

\section*{Acknowledgements}
The author gratefully acknowledge the support of the National Research Council (NCN) through the grant UMO-2013/11/B/ST3/04273 and the access to the computing facilities of the Interdisciplinary Center of Modeling at the University of Warsaw.

\section*{Author Contribution statement}
NGSz did the calculations and analysis, wrote and reviewed the manuscript, and prepared all the figures.

\section*{Competing financial interests}
The author declare no competing financial interests.

\end{document}